\documentclass[12pt,preprint]{aastex}

\slugcomment{Not to appear in Nonlearned J., 45.}

\shorttitle{\ion{He}{1} emission and absorption in DG Tau}
\shortauthors{Takami et al.}

\begin{document}


\title{\ion{He}{1} 1.083 $\mu$m emission and absorption in DG Tau: line excitation in jet, hot wind, and accretion flow\footnote{Based on data collected at Subaru Telescope, which is operated by the National Astronomical Observatory of Japan.}}


\author{M. Takami\altaffilmark{2}, A. Chrysostomou\altaffilmark{2}, J. Bailey\altaffilmark{3}, T.M. Gledhill\altaffilmark{2}, M. Tamura\altaffilmark{4}, and H. Terada\altaffilmark{5}}


\altaffiltext{2}{Department of Physical Sciences, University of Hertfordshire, College Lane, Hatfield Herts, AL10 9AB, UK} 
\altaffiltext{3}{Anglo-Australian Observatory, PO Box 296, Epping, NSW 1710, Australia} 
\altaffiltext{4}{National Astronomical Observatory of Japan, Osawa, Mitaka, Tokyo 181-8588, Japan} 
\altaffiltext{5}{Subaru Telescope, National Astronomical Observatory of Japan, 650 North A'ohoku Place, Hilo, Hawaii 96720} 
\email{takami@star.herts.ac.uk}


\begin{abstract}
We present long-slit spectroscopy and spectro-astrometry of \ion{He}{1} 1.083 $\mu$m emission in the T Tauri star, DG Tau. We identify three components in the \ion{He}{1} feature: (1) a blueshifted emission component at $v$$\simeq$$-$200 km s$^{-1}$, (2) a bright emission component at zero-velocity with a FWZI of $\sim$500 km s$^{-1}$, and (3) a blueshifted absorption feature at velocities between $-$250 and $-$500 km s$^{-1}$. The position and velocity of the blueshifted \ion{He}{1} emission coincide with a high-velocity component (HVC) of the [\ion{Fe}{2}] 1.257 $\mu$m emission, which arises from a jet within an arcsecond of the star.
The presence of such a high excitation line (excitation energy $\sim$ 20 eV) within the jet supports the scenario of shock heating. The bright \ion{He}{1} component does not show any spatial extension, and it is likely to arise from magnetospheric accretion columns.
 The blueshifted absorption shows greater velocities than that in H$\alpha$, suggesting that these absorption features arise from the accelerating wind close to the star.
\end{abstract}




\section{Introduction}
Understanding the mechanisms of mass accretion and the driving of jets/winds is one of the most important key issues of star formation theories. Disk accretion rates estimated through UV excess measurements and/or permitted line luminosities are correlated with outflow signatures such as forbidden line luminosities (e.g., Hartigan, Edwards, \& Ghandour 1995), suggesting a physical link between the two phenomena: i.e., jets/winds remove excess angular momentum from accreting material, thereby preventing young stellar objects from spinning up to breakup velocity.
Several theories have been proposed to attempt to explain this physical link (e.g., Shu et al. 2000; K\"onigl \& Pudritz 2000; Goodson, B\"ohm \& Winglee 1999), however, consensus has not yet been achieved. 

Low-mass pre-main sequence stars (T Tauri stars) exhibit a rich variety of emission lines at optical-to-IR wavelengths, probing the activities of jets, winds, and accretion flow close to the stars. 
Permitted atomic emission lines have often been considered to arise from magnetospheric accretion columns and/or accretion shocks. Indeed, these line luminosities show correlation with the accretion luminosity (e.g., Muzerolle, Hartmann, \& Carvet 1998a), and idealized magnetospheric accretion models have been successful in reproducing at least some of the observed permitted line profiles (Hartmann, Hewett, \& Carvet 1994; Muzerolle, Calvet, \& Hartmann 1998b, 2001). On the other hand,
H$\alpha$ and \ion{He}{1} profiles in some T Tauri stars exhibit blue-shifted wings, and Beristain, Edwards, \& Kwan (2001) argue that a hot wind can also contribute to their emission. Spectro-astrometric observations by Takami et al. (2001) have revealed that a bipolar outflow at AU scales is responsible for the H$\alpha$ emission in the T Tauri star RU Lupi.
Forbidden lines in T Tauri stars arise from outflowing gas at scales of more than 1 AU, and often exhibit two blueshifted components at high and low velocities (e.g., Hirth, Mundt, \& Solf 1997). Different densities and temperatures for these two components is suggestive of their distinct origins, e.g., a fast jet and a slow disk wind (see Eisl\"offel et al. 2000).

We present results of long-slit spectroscopy and spectro-astrometry for \ion{He}{1} 1.083 $\mu$m and [\ion{Fe}{2}] 1.257 $\mu$m emission in DG Tau. The former is one of the brightest among \ion{He}{1} lines at optical-to-IR wavelenths, and its diagnostic potential lies in the high-excitation energy (20 eV), which restricts its formation to a region either of high temperature or close proximity to a source of ionizing radiation (cf. Beristain et al. 2001). DG Tau is one of the most active young stellar objects known.  Analysis of the bright, hot continuum excess gives a mass accretion rate of 5$\times$10$^{-7}$ M$_{\odot}$ yr$^{-1}$ (Gullbring et al. 2000), one of the largest among T Tauri stars in the Taurus-Auriga cloud (Muzerolle et al. 1998a). Beristain et al. (2001) show a blue-shifted wing to the \ion{He}{1} 5876 {\AA} emission at velocities up to 600 km s$^{-1}$, indicating the presence of a hot wind close to the star. Recent high resolution observations have revealed knotty, bubble-like, and bow-shape structures in the associated jet within a few arcsecond of the object (e.g., Kepner et al. 1993; Dougados et al. 2000; Bacciotti et al. 2000). In this Letter, we show that the \ion{He}{1} 1.083 $\micron$ feature in DG Tau consists of the following three components: a blueshifted {\it emission} component from a jet, a blueshifted {\it absorption} component from a hot wind, and a bright emission component at zero-velocity presumably from magnetospheric accretion columns.

\section{Observations}
Observations were carried out on 2000 December 13 at the SUBARU 8.2-m
telescope using the Infrared Camera and Spectrograph IRCS (Tokunaga et al. 1998; Kobayashi et al. 2000). The echelle grating mode with a 0.3-arcsec wide slit provides a spectral resolution
of 10$^4$. The pixel scale of 0.075 arcsec provides good sampling of the seeing profile (FWHM$\sim$1 arcsec during our observations), thereby allowing accurate spectro-astrometric measurements (Bailey 1998a,b). The spectra were obtained at four slit position angles (0, 90, 180, and 270$^\circ$). The DG Tau jet lies at a position angle of $\sim 225^\circ$ (Lavalley et al. 1997; Eisl\"{o}ffel \& Mundt 1998), thus the selected slit angles minimize contamination from the extended jet and allows us to study the emission line regions closest to the star. In addition to the target, an A-type bright standard was observed at similar airmasses to correct for telluric absorption. The flat fields were made by combining many exposures of the spectrograph illuminated by a halogen lamp. 

The data were reduced using the KAPPA and FIGARO packages provided by Starlink. The position-velocity diagrams were obtained via standard reduction processes: dark-subtraction, flat-fielding, removal of bad pixels, correcting for curvature in the echelle spectra, wavelength calibration, correcting for telluric absorption, and night-sky subtraction. Wavelength calibrations were made using atmospheric absorption features on the stellar continuum via comparison with modelled spectra provided by ATRAN (Lord 1992). The systemic motion of the object was calibrated 
based on previous observations of \ion{Li}{1} 6707 {\AA} photospheric absorption (V$_{\rm{Hel}}$=+16.5 km s$^{-1}$--- Bacciotti et al. 2000).

In addition to the position-velocity diagrams, spectro-astrometric ``position spectra'' were obtained to study the spatial structure of emission line regions down to milliarcsecond scales (Bailey 1998a,b; Takami et al. 2001). These spectra were determined after dark-subtraction and flat-fielding by fitting the seeing profile at each wavelength with a Gaussian function.
About 6$\times$10$^4$ photons at each wavelength provide a typical accuracy of 6 milliarcsec at the continuum level.
Any instrumental effects in the position spectra were eliminated with high accuracy by subtracting those with opposite position angles ($0 - 180 ^\circ$ or $90 - 270 ^\circ$).
Each position spectrum has an arbitrary zero point, which is adjusted to correspond to the continuum position. To determine the true-position of the line emission, contamination by the continuum was removed using the line-to-continuum ratio at each wavelength (see Takami et al. 2001 for detail).

\section{Results}
Fig. 1 shows the position-velocity diagrams for \ion{He}{1} 1.083 $\mu$m and [\ion{Fe}{2}] 1.257 $\mu$m emission along north-south (N-S) and east-west (E-W) slit positions.
The figure exhibits two components in both the \ion{He}{1} and [\ion{Fe}{2}] emission.
The \ion{He}{1} emission consists of a bright component centered on the systemic velocity, and a high velocity component (HVC) at a velocity of roughly $-200$ km s$^{-1}$.
The former component does not show any spatial displacement from the star, whose position is indicated by the dot-dashed line in the figure.
On the other hand, the HVC extends towards the south and west peaking 0".4 south of the star.
The position of the HVC coincides with structure in the jet located $\sim$0".6 southwest from the star, previously identified by Bacciotti et al. (2000).

The position-velocity diagrams of the [\ion{Fe}{2}] emission exhibit two blueshifted velocity components as observed in optical forbidden lines of many T Tauri stars (e.g., Hirth et al. 1997).
These components in our results exhibit velocities of $-$200 to $-$250 km s$^{-1}$ and $-$100 to $-$150 km s$^{-1}$, and their centroidal positions are displaced from the star in the direction of the jet by $\sim$0".4 arcsec and $\sim$0".2 arcsec, respectively, in both the N-S and E-W diagrams. 
The position and the velocity of the HVC in the [\ion{Fe}{2}] emission coincide with those of the \ion{He}{1} emission described above. 

Fig. 2 shows intensity profiles in the regions (a)(b)(c) given in Fig. 1: each 0", 0".4, and 0".8 arcsecs from the star in the north-south diagrams, respectively. The \ion{He}{1} 1.083 $\mu$m profile in region (a) exhibits a triangular component at zero velocity with a FWZI of $\sim$500 km s$^{-1}$, corresponding to the bright component in Fig. 1.
The HVC clearly appears in profiles (b) and (c), exhibiting a peak at $-$205 km s$^{-1}$.
In addition to these emission components, profile (a) shows blueshifted absorption between $-$250 and $-$500 km s$^{-1}$.
The intensity profiles in [\ion{Fe}{2}] emission in Fig. 2 exhibit blueshifted peaks at $-$230 to $-$240 and $-$105 km s$^{-1}$, respectively. These [\ion{Fe}{2}] profiles are similar to those of optical emission lines observed by Hamann (1994), especially of [\ion{Fe}{2}] 8617 {\AA}, 7452 {\AA}, and 7388 {\AA} lines.

Fig. 2 also shows position spectra obtained via spectro-astrometric reduction. Displacement is not detected in the brightest emission component of the \ion{He}{1} emission within the measured positional accuracy of $\sim$10 milliarcsec, corresponding to $\sim$1.4 AU at the observed star-forming region (Wichmann et al. 1998). This is consistent with the argument that permitted atomic emission lines in T Tauri stars form within $\sim$10 R$_*$ of the stellar surface (e.g., Najita et al. 2000). The position spectra in the low-velocity component (LVC) of the [\ion{Fe}{2}] emission show a constant displacement of $\sim$0".2 within the error bars. A constant displacement is also observed in the optical forbidden lines of RU Lupi (Takami et al. 2001), and contrasts with that expected in an accelerating flow in which the displacement should increase with the velocity. Such constancy in the position spectra could be explained if the line broadening is due to rotation around the flow axis (Hirth, Mundt, \& Solf 1994), although the error bars in Fig. 2 cannot rule out the contribution of accelerating motion. 
The measured position angles in the emission features are $215\pm 6 ^\circ$, $213 \pm 14 ^\circ$, and $219\pm 17 ^\circ$ for the \ion{He}{1} HVC, [\ion{Fe}{2}] HVC, and [\ion{Fe}{2}] LVC, respectively, coincident with the position angle of the extended jet (Lavalley et al. 1997; Eisl\"{o}ffel and Mundt 1998).

\section{Discussion}
It is surprising that the \ion{He}{1} 1.083 $\mu$m emission is detected within the HVC of the forbidden line region. The HVC of forbidden line regions in T Tauri stars have been considered to arise from a jet close to the source (Eisl\"offel et al. 2000), and mainly observed in [\ion{O}{1}] 6300 {\AA}, [\ion{S}{2}] 6716/6731 {\AA}, and [\ion{N}{2}] 6583 {\AA}, whose excitation energies are typically 2 eV from the ground. On the other hand, the \ion{He}{1} 1.083 $\mu$m emission has an excitation energy of 20 eV, about $\sim$10 times larger. This line has been observed in some high-velocity Harbig-Haro objects (e.g., Brugel, B\"ohm, \& Mannery 1981) and in \ion{H}{2} regions (e.g., Takami et al. 2002), however, it is not easily thermally excited, requiring conditions with temperatures of $\sim$10$^4$ K and electron densities of $<$10$^6$ cm$^{-3}$.

The heating mechanism in the jet close to the source has been debated over many years. Proposed heating mechanisms include: shocks caused by variations in the velocity of the jet (e.g., Hartigan et al. 1995), ambipolar diffusion (Safier 1993; Garcia 2001), and turbulent dissipation in a viscous mixing-layer (e.g., Binette et al. 1999). Recent high-resolution imaging and emission line diagnostics have prefered shock heating (Bacciotti et al. 2000; Lavalley-Fouquet, Cabrit, \& Dougados 2000), and our results support this conclusion for the following reasons. One of the outer working surfaces of the DG Tau jet exhibit [\ion{O}{3}] 4959 {\AA} and 5007 {\AA} emission (Cohen \& Fuller 1985), indicating that O$^+$ ions are ionized beyond their ionization potential of 35 eV: thus, shock velocities in the DG Tau jet should be sufficient to excite the \ion{He}{1} 1.083 $\mu$m emission, the excitation energy of which is 20 eV.
On the other hand, the models of ambipolar diffusion and turbulent dissipation predict temperatures of up to $\sim$$3 \times 10 ^4$ K (Safier 1993; Garcia et al. 2001; Binette et al. 1999) which, together with the inferred electron densities ($\sim$10$^5$ cm$^{-3}$ --- Lavalley-Fouquet et al. 2000), is not likely to explain the presence of such a highly excited line.

The velocities of the \ion{He}{1} blueshifted $absorption$ roughly coincide with the blueshifted wing $emission$ of \ion{He}{1} 5576 {\AA}, which extends up to 600 km s$^{-1}$ (Beristain et al. 2001). This difference is attributed to their different optical thickness. The lower energy level of the \ion{He}{1} 1.083 $\mu$m transition is ``metastable'', and its large population causes an absorption feature via resonance scattering in the outer region of a non-spherical wind. On the other hand, the lower level of the \ion{He}{1} 5576 {\AA} transition must be much less populated, allowing the emission from the inner wind to pass through the outer region without absorption. 
Our results do not show the presence of off-stellar \ion{He}{1} emission associated with the blueshifted absorption, as observed in a massive star $\theta ^1$C Ori (Oudmaijer et al. 1997). This fact suggest that the \ion{He}{1} wind in DG Tau has a spatial scale well below our spatial resolution. Interestingly, the H$\alpha$ profiles observed by Beristain et al. (2001) exhibit a blueshifted absorption at much lower velocities, peaking at roughly $-$50 km s$^{-1}$. Such different velocities between the \ion{He}{1} 1.083 $\mu$m and H$\alpha$ absorption may be a result of acceleration in the wind. 

The bright \ion{He}{1} emission component at zero velocity appears both in \ion{He}{1} 1.083 $\mu$m and 5876 {\AA}, in contrast with the wind at $-250$ to $-500$ km s$^{-1}$. Such different line excitation between the two components suggest distinct origins. A possible explanation is that the bright component arises in magnetospheric accretion columns, as often argued for permitted lines. Indeed, the profiles of this component are similar to those of Pa$\beta$ (Folha \& Emerson 2001), whose luminosity show a tight correlation with the accretion luminosity (Muzerolle et al. 1998a). 


\vspace{0.5cm}
The authors thank the anonymous referee for useful comments, and T. Onaka for use of ATRAN on his workstation.
The authors also acknowledge the data analysis facilities provided by the
Starlink Project which is run by CCLRC on behalf of PPARC.  MT thanks PPARC for support through a PDRA.
 



\clearpage


\begin{figure*}
\plotone{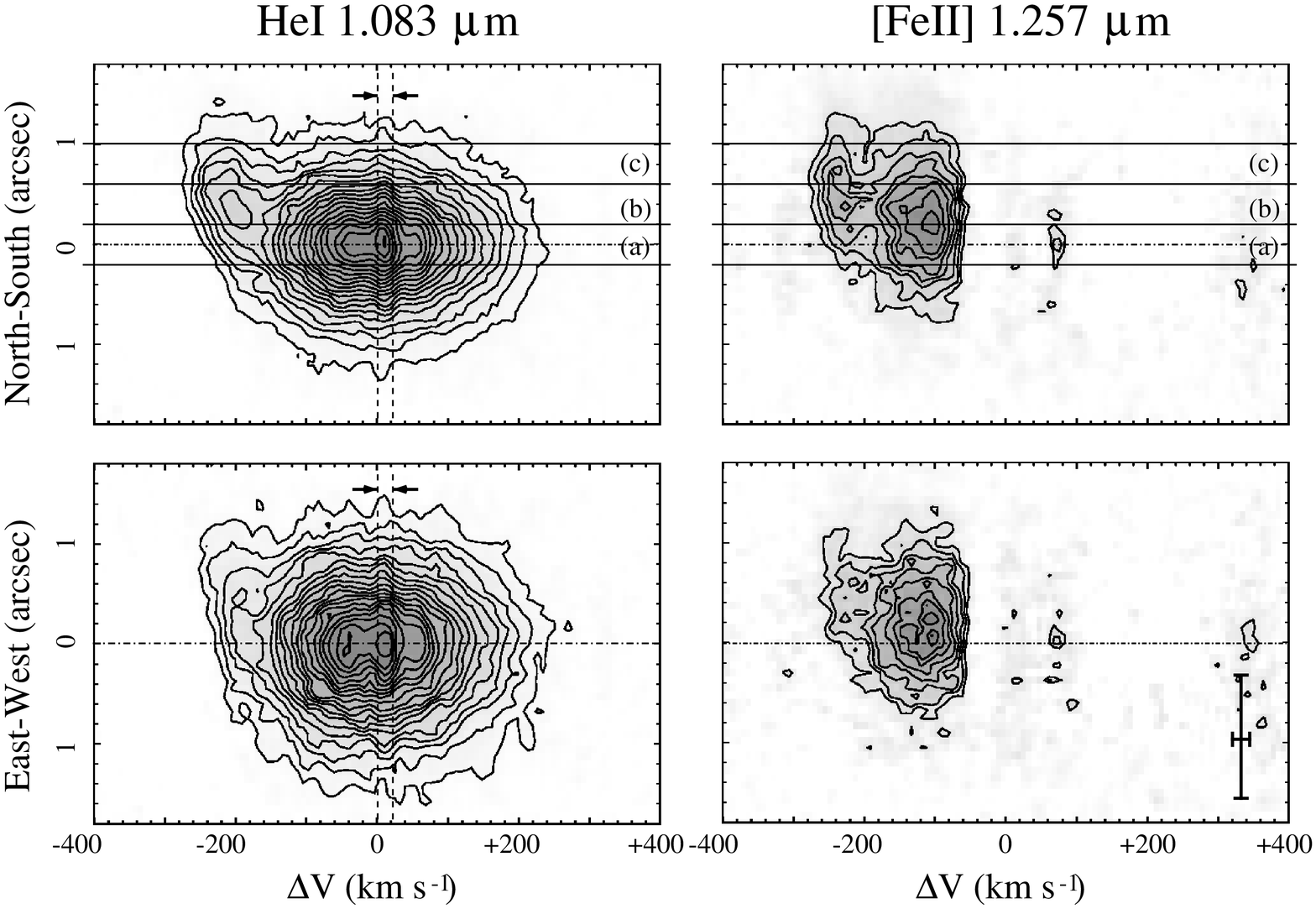}
\caption{Continuum-subtracted position-velocity diagrams for the \ion{He}{1} 1.083 $\mu$m and [\ion{Fe}{2}] 1.257$\mu$m lines in the north-south and east-west directions. The spacing of the contour corresponds to 5\% and 10\% of the peak flux in the \ion{He}{1} and [\ion{Fe}{2}] diagrams, respectively.
A dot-dashed line in each figure shows the position of the star. Scale bars in the bottom-right figure show the average seeing size and the spectral resolution. Intensity profiles are extracted from the regions (a)(b)(c), and shown in Figure 2.
Excess emission between dotted lines may be due to imperfect sky subtraction. \label{fig1}}
\end{figure*}

\clearpage 

\begin{figure*}
\plotone{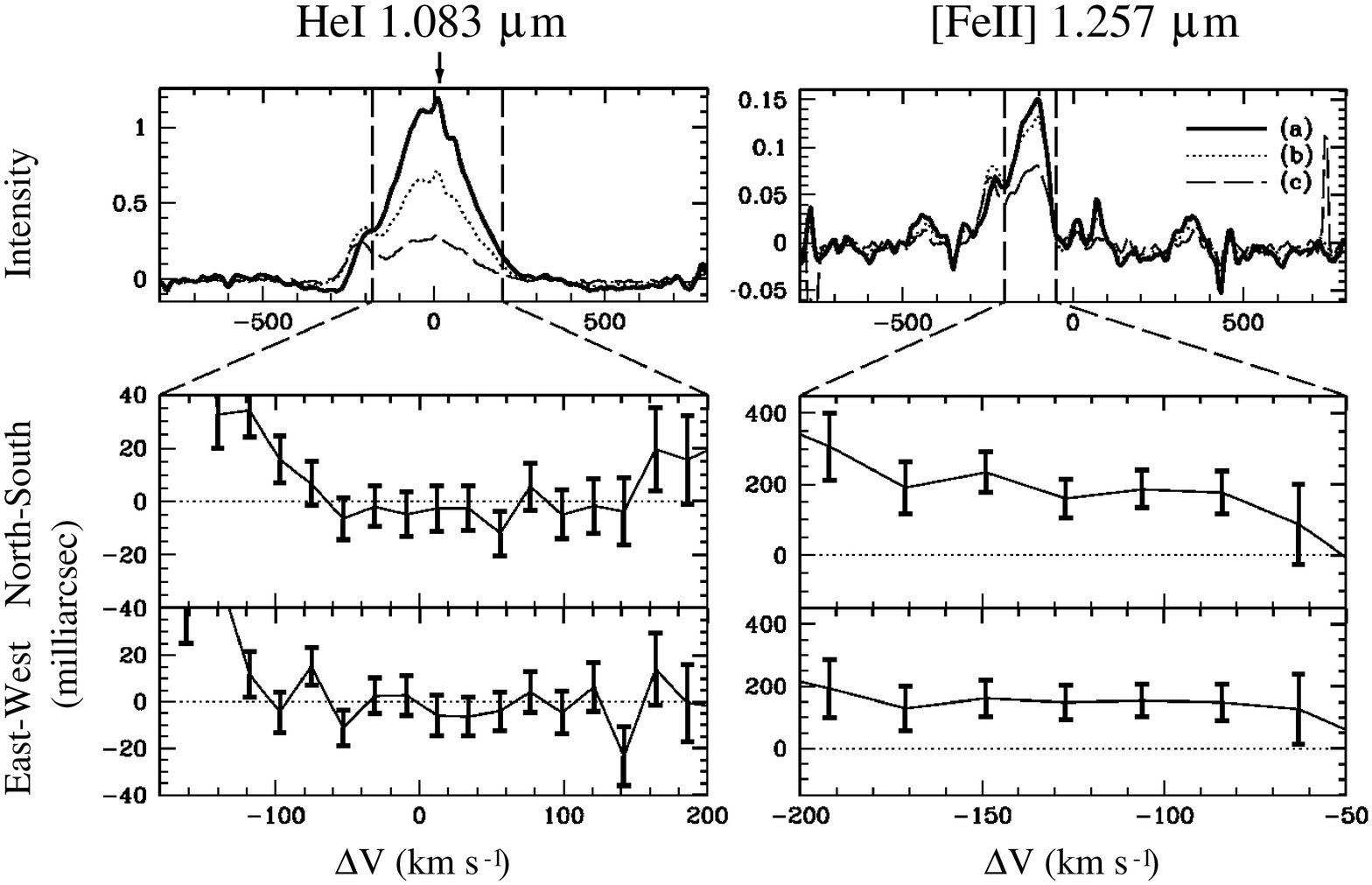}
\caption{Intensity and position spectra of the \ion{He}{1} 1.083 $\mu$m and [\ion{Fe}{2}] 1.257$\mu$m lines. Solid, dotted, and dashed intensity profiles were obtained from the regions (a)(b)(c) in the N-S diagrams (see Fig. 1), and normalized to the continuum flux in (a). Position spectra were obtained from the whole seeing profile including these three regions. The narrow hump in the \ion{He}{1} profiles indicated by the arrow may be due to imperfect sky subtraction. Position spectra are shown in the velocity ranges in which spatial structure is marginally or negatively shown in Fig. 1. \label{fig2}}
\end{figure*}


\begin{thebibliography}{}
\bibitem[Bailey(1998a)]{1998SPIE.3355..932B} Bailey, J.~A.\ 1998, \procspie, 3355, 932
\bibitem[Bailey(1998b)]{1998MNRAS.301..161B} Bailey, J.\ 1998, \mnras, 301, 161
\bibitem[Bacciotti et al.(2000)]{2000ApJ...537L..49B} Bacciotti, F., Mundt, R., Ray, T.~P., Eisl{\" o}ffel, J., Solf, J., \& Camezind, M.\ 2000, \apjl, 537, L49
\bibitem[Beristain et al.~2001]{BEK00} Beristain, G., Edwards, S., \& Kwan, J.\ 2001, \apj,  551, 1037 
\bibitem[Binette et al.~1999]{Binette99} Binette, L., Cabrit, S., Raga, A., \& Cant{\' o}, J.\ 1999, \aap,  346, 260
\bibitem[Brugel, Boehm, \& Mannery(1981)]{1981ApJS...47..117B} Brugel, E.~W., B\"ohm, K.~H., \& Mannery, E.\ 1981, \apjs, 47, 117
\bibitem[Cohen \& Fuller 1985]{Cohen85} Cohen, M. \& Fuller, G.~A.\ 1985, \apj,  296, 620
\bibitem[Dougados, Cabrit, Lavalley, \& M{\' e}nard(2000)]{2000A&A...357L..61D} Dougados, C., Cabrit, S., Lavalley, C., \& M{\' e}nard, F.\ 2000, \aap, 357, L61
\bibitem[Eisl{\" o}ffel \& Mundt(1998)]{1998AJ....115.1554E} Eisl{\" o}ffel, J.~\& Mundt, R.\ 1998, \aj, 115, 1554
\bibitem[Eis\"offel, Mundt, Ray, \& Rodriguez(2000)]{2000prpl.conf..815E} Eis\"offel, J., Mundt, R., Ray, T.~P., \& Rodriguez, L.~F.\ 2000, Protostars and Planets IV (Book - Tucson: University of Arizona Press; eds Mannings, V., Boss, A.P., Russell, S.~S.), p.~815, 815
\bibitem[Folha \& Emerson 2001]{2001A&A...365...90F} Folha, D.~F.~M.~\& Emerson, J.~P.\ 2001, \aap,  365, 90
\bibitem[Garcia et al.~2001a]{Garcia01} Garcia, P.~J.~V., Ferreira, J., Cabrit, S., \& Binette L.\ 2001, \aap, 377, 589
\bibitem[Goodson et al.~1999]{G1999} Goodson, A.~P., B{\" o}hm, K., \& Winglee, R.~M.\ 1999, \apj,  524, 142
\bibitem[Gullbring, Calvet, Muzerolle, \& Hartmann(2000)]{2000ApJ...544..927G} Gullbring, E., Calvet, N., Muzerolle, J., \& Hartmann, L.\ 2000, \apj, 544, 927
\bibitem[Hamann 1994]{H94} Hamann, F.\ 1994, \apjs, 93, 485
\bibitem[Hartigan et al.~1995]{HEG95} Hartigan, P., Edwards, S., \& Ghandour, L.\ 1995, \apj,  452, 736
\bibitem[Hartmann et al.~1994]{H94} Hartmann, L., Hewett, R., \& Calvet, N.\ 1994, \apj,  426, 669
\bibitem[Hirth, Mundt, \& Solf(1997)]{1997A&AS..126..437H} Hirth, G.~A., Mundt, R., \& Solf, J.\ 1997, \aaps, 126, 437
\bibitem[Kepner et al.~1993]{Kepner93} Kepner, J., Hartigan, P., Yang, C., \& Strom, S., 1993, \apjl,  415, L119
\bibitem[Kobayashi et al.(2000)]{2000SPIE.4008.1056K} Kobayashi, N.~et al.\ 2000, \procspie, 4008, 1056
\bibitem[Konigl \& Pudritz(2000)]{2000prpl.conf..759K} Konigl, A.~\& Pudritz, R.~E.\ 2000, Protostars and Planets IV (Book - Tucson: University of Arizona Press; eds Mannings, V., Boss, A.P., Russell, S.~S.), p.~759, 759
\bibitem[Lavalley et al.(1997)]{1997A&A...327..671L} Lavalley, C., Cabrit, S., Dougados, C., Ferruit, P., \& Bacon, R.\ 1997, \aap, 327, 671
\bibitem[Lavalley-Fouquet, Cabrit, \& Dougados(2000)]{2000A&A...356L..41L} Lavalley-Fouquet, C., Cabrit, S., \& Dougados, C.\ 2000, \aap, 356, L41
\bibitem[Lord(1992)]{lor92} Lord, S. D.\ 1992, NASA Technical Memor. 103957 
\bibitem[Muzerolle et al.~1998]{Muzerolle98a} Muzerolle, J., Calvet, N., \& Hartmann, L.\ 1998b, \apj,  492, 743
\bibitem[Muzerolle et al.~2001]{Muzerolle01} Muzerolle, J., Calvet, N., \& Hartmann, L.\ 2001, \apj,  550, 944
\bibitem[Muzerolle et al.~1998]{Muzerolle98b} Muzerolle, J., Hartmann, L., \& Calvet, N., 1998a, \aj,  116, 2965
\bibitem[Najita et al.~2000]{N2000} Najita, J.~R., Edwards, S., Basri, G., \& Carr J., 2000  Protostars and Planets IV (Book - Tucson: University of Arizona Press; eds Mannings, V., Boss, A.P., Russell, S.~S.), p.~457
\bibitem[Oudmaijer, Busfield, \& Drew(1997)]{1997MNRAS.291..797O} Oudmaijer, R.~D., Busfield, G., \& Drew, J.~E.\ 1997, \mnras, 291, 797
\bibitem[Safier 1993a]{Safier93a} Safier, P.~N.\ 1993, \apj,  408, 115
\bibitem[Shu et al.~2000]{Shu00} Shu, F.~H., Najita, J.~R., Shang, H., \& Li, Z.-Y., 2000\ , Protostars and Planets IV (Book - Tucson: University of Arizona Press; eds Mannings, V., Boss, A.P., Russell, S.~S.), p.~789
\bibitem[Takami et al.(2001)]{2001MNRAS.323..177T} Takami, M., Bailey, J., Gledhill, T.~M., Chrysostomou, A., \& Hough, J.~H.\ 2001, \mnras, 323, 177
\bibitem[Takami et al.~2002]{Takami02} Takami, M. et al. 2002, \apj, 566, 910
\bibitem[Tokunaga et al.(1998)]{1998SPIE.3354..512T} Tokunaga, A.~T.~et al.\ 1998, \procspie, 3354, 512
\bibitem[Wichmann et al.(1998)]{1998MNRAS.301L..39W} Wichmann, R., Bastian, U., Krautter, J., Jankovics, I., \& Rucinski, S.~M.\ 1998, \mnras, 301, L39
\end{thebibliography}
\end{document}